# A Scalable and Quantum-Accurate Foundation Model for Biomolecular Force Field via Linearly Tensorized Quadrangle Attention

Qun Su[1, #], Kai Zhu[1, #], Qiaolin Gou[2, #], Jintu Zhang[1], Renling Hu[1], Yurong Li[1], Yongze Wang[1], Hui Zhang[1], Ziyi You[1], Linlong Jiang[1], Yu Kang[1], Jike Wang[1, *], Chang-Yu Hsieh[1, *], Tingjun Hou[1, *]

[1]*College of Pharmaceutical Sciences, Zhejiang University, Hangzhou 310058, Zhejiang, China*

[2]*Faculty of Applied Sciences, Macao Polytechnic University, Macao 999078, SAR, China.*

[#]*Equivalent authors*

## Corresponding authors

**Tingjun Hou**
**E-mail:** tingjunhou@zju.edu.cn
**Chang-Yu Hsieh**
**E-mail:** kimhsieh@zju.edu.cn
**Jike Wang**
**E-mail:** jikewang@zju.edu.cn

# Abstract

Accurate atomistic biomolecular simulations are vital for disease mechanism understanding, drug discovery, and biomaterial design, but existing simulation methods exhibit significant limitations. Classical force fields are efficient but lack accuracy for transition states and fine conformational details critical in many chemical and biological processes. Quantum Mechanics (QM) methods are highly accurate but computationally infeasible for large-scale or long-time simulations. AI-based force fields (AIFFs) aim to achieve QM-level accuracy with efficiency but struggle to balance many-body modeling complexity, accuracy, and speed, often constrained by limited training data and insufficient validation for generalizability. To overcome these challenges, we introduce LiTEN, a novel equivariant neural network with Tensorized Quadrangle Attention (TQA). TQA efficiently models three- and four-body interactions with linear complexity by reparameterizing high-order tensor features via vector operations, avoiding costly spherical harmonics. Building on LiTEN, LiTEN-FF is a robust AIFF foundation model, pre-trained on the extensive nablaDFT dataset for broad chemical generalization and fine-tuned on SPICE for accurate solvated system simulations. LiTEN achieves state-of-the-art (SOTA) performance across most evaluation subsets of rMD17, MD22, and Chignolin, outperforming leading models such as MACE, NequIP, and EquiFormer. LiTEN-FF enables the most comprehensive suite of downstream biomolecular modeling tasks to date, including QM-level conformer searches, geometry optimization, and free energy surface construction, while offering 10× faster inference than MACE-OFF for large biomolecules (~1000 atoms). In summary, we present a physically grounded, highly efficient framework that advances complex biomolecular modeling, providing a versatile foundation for drug discovery and related applications.

# Introduction

Accurate atomistic simulations are essential for understanding disease mechanisms, advancing drug discovery, and designing novel biomaterials by elucidating biomolecular structures and functions[1]. Traditional methods fall into two categories: classical force fields and quantum chemical calculations. Classical force fields enable large-scale molecular dynamics (MD) simulations due to their high computational efficiency but struggle to accurately model bond rearrangements, transition state energetics, and fine conformational nuances due to fixed-parameters , often leading to discrepancies in dynamic simulations[2]. In contrast, quantum chemical methods (e.g., density functional theory) explicitly describe electronic structures and bond transformations, making them invaluable for reaction pathway analysis and small molecule optimization[3, 4]. However, their high computational cost restricts their use to short-timescale or small-system simulations, limiting their applicability to complex biomolecular dynamics. Thus, achieving a practical balance between physical accuracy and computational efficiency remains a fundamental and ongoing challenge in biomolecular modeling[5, 6].

Recent advances in deep learning have opened new venues for addressing challenges in biomolecular modeling. Artificial Intelligence Force Fields (AIFFs), trained on quantum chemical data, have demonstrated a favorable trade-off between accuracy and computational efficiency, enabling physically grounded modeling of complex molecular systems[7-9]. The evolution of AIFFs follows three key trends: (1) transitioning from global molecular descriptors to localized atomic environment representations; (2) advancing from approximate symmetry handling to rigorous physical equivariances; and (3) extending from simple two-body to comprehensive many-body interactions[10, 11]. This progression has given rise to several influential methodological frameworks. SchNet[12] pioneered continuous-filter convolutional networks for end-to-end prediction of molecular energies and forces. PaiNN[13] incorporated rotation-equivariant message passing, significantly enhancing the model's capacity to encode molecular symmetries. NequIP[14] and MACE[15] developed graph neural network architectures that strictly comply with E(3) equivariance, employing high-order tensor products to model many-body interactions, achieving state-of-the-art (SOTA) performance on small molecules. AIMNet2[16] proposed hierarchical multi-scale interaction modules to capture long-range electronic polarization effects with robust generalizability. The ANI[17] series focused on drug-like molecules, combining local atomic environment descriptors with large-scale training datasets for practical drug screening workflows.

Despite the promise of AIFFs in various applications, critical challenges remain in biomolecular modeling. First, while datasets have expanded in scale and coverage, they are fragmented and rely on ab initio methods with varying accuracy (e.g., DFT[18], MP2[19], CCSD(T)[20]), lacking a unified computational benchmark[21, 22]. This methodological heterogeneity introduces biases in potential energy surfaces (PES), hindering cross-dataset training and transfer learning. Second, while datasets such as QM9[23], ANI-1x, and AIMNet2 cover the general organic chemical space, their representation of key biomolecular subspaces (e.g., carbohydrates, nucleotides, metal complexes, and polar residues) remains insufficient, limiting model generalizability and predictive stability in real biological systems.

Given the limitations in existing datasets, current models often struggle with limited generalization and reduced prediction accuracy. To improve performance, many approaches have adopted more complex neural network architectures, which significantly increase computational inference cost[24]. High-order equivariant models (e.g., NequIP and MACE) demonstrate superior physical consistency and many-body modeling but suffer from low inference efficiency due to complex construction and propagation of high-order tensors, especially for medium-to-large systems. In contrast, lower-order models like AIMNet2, built upon Cartesian coordinates without explicit tensor products, achieve faster inference speed, making them more suitable for industrial deployment. However, they sacrifice predictive accuracy for non-local coupling or large systems, as their inability to represent high-order tensors restricts capturing subtle many-body effects critical for local potential energy surface curvature and coupling directions. Therefore, in the design of AIFFs, completely omitting tensor product operations within the Cartesian coordinate framework can substantially boost computational efficiency, though this comes at the cost of enforcing symmetry constraints and compromising high-order many-body modeling. Conversely, employing high-order tensor products enhances representational power but incurs significant computational costs, limiting practicality for large molecular systems. Consequently, there is an urgent need for a novel framework that balances physical consistency, modeling accuracy, and computational efficiency, alongside new solutions for pretraining data and strategies.

In this study, we proposed LiTEN, an equivariant tensor message-passing model built in Cartesian coordinates that combines physical consistency, expressive power, and inference efficiency. Its core innovation is the Tensorized Quadrangle Attention (TQA) mechanism, which unifies the modeling of three-body and four-body interactions centered on edges and nodes with linear complexity. This enables efficient representation of many-body couplings in complex molecular conformations. Unlike traditional equivariant networks that rely on explicit high-order tensor products and spherical harmonic expansions, LiTEN reparameterizes high-order tensor structures using dot and cross products between vectors, compactly encoding interactions while preserving equivariance. This approach significantly reduces computational graph complexity and circumvents the performance bottleneck associated with Clebsch-Gordan coefficient calculations. This design not only improves the model's structural awareness in dense many-body fields but also offers a physically motivated and scalable solution for AI-driven force field modeling in bioorganic systems. Evaluations on multiple benchmark datasets, including rMD17[25], MD22[26] and Chignolin[27], demonstrate that LiTEN achieves SOTA performance across most test subsets, outperforming established methods such as MACE, NequIP, and Allegro[28].

Built on the LiTEN architecture, we introduce LiTEN-FF, a robust foundation model for atomistic simulation. To balance quantum-level accuracy and broad chemical generalization, LiTEN-FF is trained through a two-stage strategy: pretraining on the large-scale nablaDFT[29] dataset (~16 million conformations of drug-like molecules with elements H, C, N, O, S, Cl, F, Br), followed by fine-tuning on the higher-accuracy SPICE[30] dataset (~2 million bio-organic molecules covering a wider element range including metals and halogens). This hierarchical approach equips LiTEN-FF with strong transferability across diverse chemical and biological systems, including solvated environments. In terms of modeling capability, LiTEN-FF achieves quantum accuracy in conformer optimization across systems of varying sizes, with an average RMSD of just 0.048 Å compared to DFT references. In MD, it reproduces bond length and angle distributions with KL divergences as low as 0.001–0.01. On the TorsionNet206[31] benchmark, it outperforms existing AIFFs and medium-level DFT methods, achieving a MAE of only 0.19 kcal/mol relative to CCSD(T). Its predicted free energy surfaces closely match experimental observations, confirming its suitability for thermodynamic modeling. In periodic aqueous simulations, LiTEN-FF accurately reproduces radial distribution functions and performs similarly to MACE-OFF[32] in modeling peptide free energy landscapes. For computational performance, LiTEN-FF delivers up to a 10× speedup over leading models like MACE-OFF on large systems (~1000 atoms), while maintaining or exceeding their accuracy. It also achieves the first batch-mode conformer search among AIFFs, scaling up to 40× faster with increased batch size—highlighting its strength in high-throughput molecular modeling.

In summary, LiTEN and LiTEN-FF overcome key limitations of existing force fields, delivering a scalable, quantum-accurate solution that pushes the boundaries of AIFFs in both speed and accuracy. Their integration with structure prediction and molecular design pipelines promises to accelerate real-world applications in drug discovery and materials science.

## Results and Discussion

### Model Architecture

LiTEN adopts a hierarchical architecture **(Figure 1A-1B)**, consisting of an embedding layer designed to capture local atomic geometric features, followed by three core modules responsible for tensor-based message passing, edge-based vector cross-product interactions, and node-based vector-scalar fusion. The architecture culminates with output heads that predict total molecular energy and atomic forces. Importantly, the atomic forces are computed as the negative gradient[36] of the energy with respect to atomic coordinates, ensuring physical consistency and stability for MD simulations.

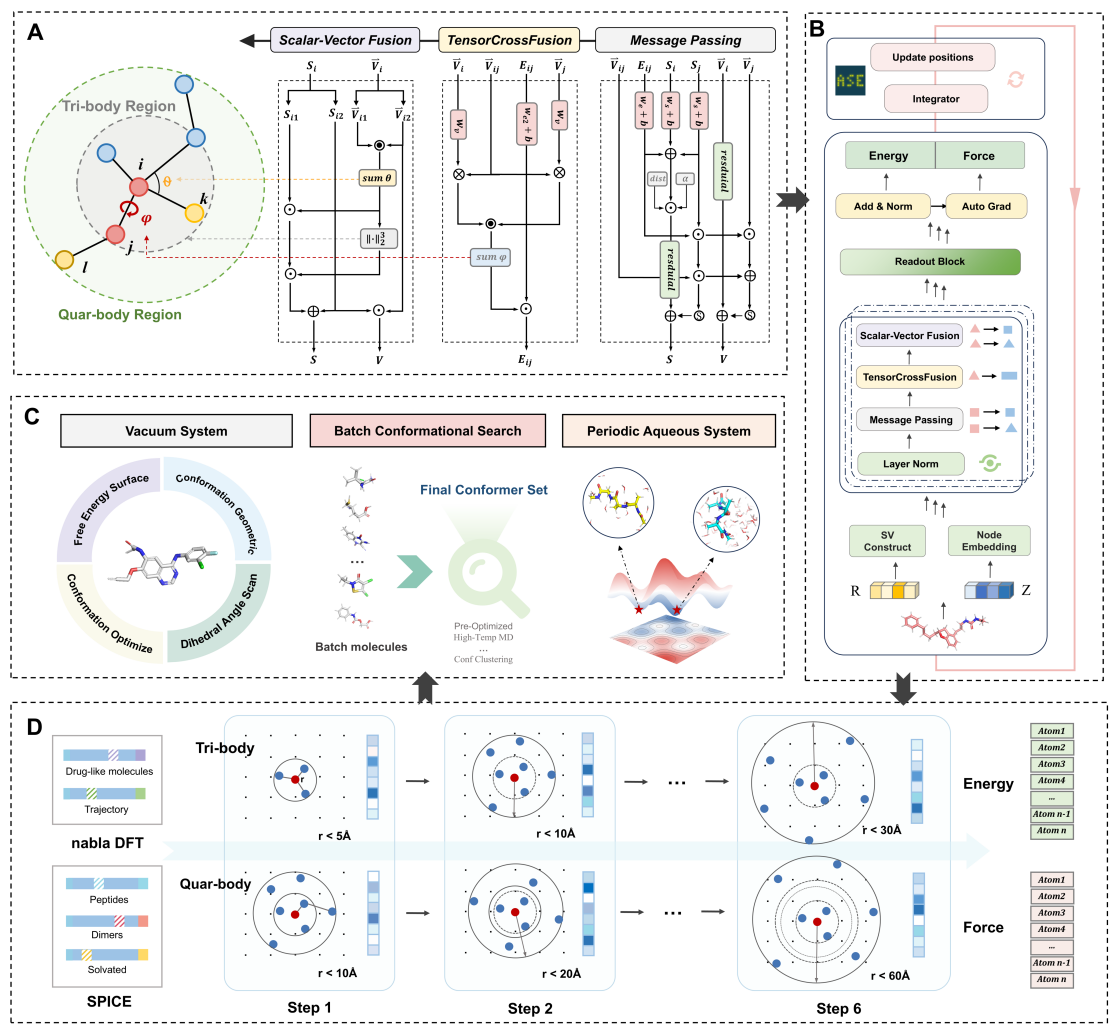

**Figure 1.** The workflow of LiTEN-FF. (A) The core architecture of LiTEN, illustrating the geometric computations and their corresponding physical interpretations in terms of multi-body interactions. The operations are denoted as follows: ⊙ represents the Hadamard product, ⊗ the cross product, ⊕ summation, and the final symbol indicates the dot product. (B) The overall pipeline of LiTEN is compatible with simulation frameworks such as ASE[33], enabling downstream applications including molecular dynamics and geometry optimization after training. (C) The foundation model LiTEN-FF can be applied to a variety of scenarios under both vacuum and solvated conditions, supporting tasks such as conformer search, free energy surface construction, dihedral angle scans, and geometry evaluation. (D) Hierarchical two-stage training is conducted using large-scale quantum chemical datasets including nablaDFT and SPICE. The model contains six interaction layers. In each layer, the receptive field for three-body interactions is 5 Å, while four-body interactions extend to 10 Å. As depth increases, the receptive field grows up to 60 Å, allowing effective modeling of both short- and long-

range interactions.

In the model design, each atom $i$ is assigned a first-order equivariant vector $\vec{u_i}$, Given the neighbor set $\mathcal{N}(i)$ centered on atom $i$, we perform a weighted aggregation of the normalized displacement vectors $\vec{u_{ij}}$ from all neighbors to obtain a first-order geometric descriptor:

$$\vec{u_i} = \sum_{j \in \mathcal{N}(i)} \vec{u_{ij}}, \quad \vec{u_{ij}} = \frac{\vec{r_{ij}}}{|\vec{r_{ij}}|}. \tag{1}$$

**Node-based many-body interactions**: This vector $\vec{u_i}$ encodes the dominant directional characteristics (i.e., first-order anisotropy) of the atom within its local environment. Based on the inner product of directional vectors, a three-body structural embedding can be constructed without explicit angle calculations:

$$|\vec{u_i}|^2 = \sum_{j,k \in \mathcal{N}(i)} \vec{u_{ij}} \cdot \vec{u_{ik}} = \sum_{j,k} \cos \theta_{jik}. \tag{2}$$

This quadratic form is equivalent to the sum of the cosines of the angles between adjacent edges. When neighboring atoms tend to align collinearly, this value reaches its maximum, reflecting the angular consistency of the atom's local environment. To better distinguish conformations with different symmetries or principal axis distributions, we introduce a cubic composite tensor with two main purposes. Firstly, it aims to embed angular information along with the magnitudes of atomic vectors into the node scalars, thereby enhancing the model's sensitivity to directional shifts in the local conformation. Its expression is given by:

$$|\vec{u_i}|^3 = \left( \sum \cos \theta_{jik} \right) \cdot |\vec{u_i}|. \tag{3}$$

Secondly, this cubic term integrates both self-interaction and cross-interaction information, effectively capturing weakly coupled many-body structures. Its approximate expansion is as follows:

$$|\vec{u_i}|^3 = (A + B)^{3/2} \approx A^{3/2} + \frac{3}{2} A^{1/2} B + \frac{3}{8} A^{-1/2} B^2. \tag{4}$$

Here, we define $A = \sum_j |\vec{u_{ij}}|^2$ as the self-interaction term, and $B = \sum_{j \neq k} \vec{u_{ij}} \cdot \vec{u_{ik}}$ as the cross-interaction term. The above formula is a Taylor expansion valid under the condition $|B/A| \ll 1$. This approximation relies on the selection of neighboring atoms.

Typically, local neighbors are obtained within a cutoff radius, which comprehensively covers the atomic distribution around the target atom. In the ideal case of a uniform atomic distribution, $B \approx 0$, and in other cases, $B$ generally remains small, effectively describing weakly coupled many-body interactions[37, 38]. When the local atomic arrangement exhibits a distinct directional bias (e.g., linear or L-shaped conformations), this term's response is enhanced, whereas it approaches zero in perfectly symmetric conformations (e.g., tetrahedral), thereby improving the distinguishability of directional structures. These invariants, constructed from directional tensors, collectively form a multibody descriptive basis that does not require manual construction of angular terms.

**Edge-based four-body interactions**: To model higher-order conformational constraints, such as torsional potentials and spatial rigidity, we propose an edge-oriented four-body interaction mechanism. For a quadruple $(i, j, k, l)$, where each pair of atoms is connected through shared edges, forming a chain-like structure, we define the cosine of the dihedral (torsion) angle between adjacent edges as follows:

$$\cos \phi = \frac{(\vec{u_{ik}} \times \vec{u_{ij}}) \cdot (\vec{u_{ij}} \times \vec{u_{jl}})}{|\vec{u_{ik}} \times \vec{u_{ij}}| \cdot |\vec{u_{ij}} \times \vec{u_{jl}}|} \tag{5}$$

This formula is equivalent to the cosine of the angle between the normal vectors of two adjacent planes, capturing the spatial torsion information present in the four-atom chain.

By incorporating the sine of the angle, $\sin \theta$, and considering that the angular component $\cos \theta$ is already captured in the three-body interactions, we construct the following rotation-invariant four-body geometric descriptor:

$$\sum_{k,l \in \mathcal{N}(i,j)} \sin \theta_{kij} \sin \theta_{lji} \cos \phi = \sum_{k,l} (\vec{u_{ik}} \times \vec{u_{ij}}) \cdot (\vec{u_{ij}} \times \vec{u_{jl}}) = (\vec{u_i} \times \vec{u_{ij}}) \cdot (\vec{u_{ij}} \times \vec{u_j}) \tag{6}$$

This information is encoded into edge features within the model and used for calculating attention weights in subsequent iterations. Specifically, the weights in the implicit attention mechanism[39] are computed by directly summing the embeddings of nodes and edges, followed by applying activation and scaling, and then combining with a distance decay factor:

$$\alpha_{ij} = [\phi(h_i + h_j + e_{ij}) \cdot \alpha]^\top \cdot R(d_{ij}) \qquad (7)$$

This component enhances the model's sensitivity to molecular conformational distortions, rigid fragments, and deviations from coplanarity, making it well-suited for accurately characterizing complex structures such as conjugated systems, helical chains, and protein backbones. It is worth noting that VisNet constructs dihedral angles via the vector rejection method. Although this approach is effective, it integrates surrounding vectors without explicitly normalizing the rejection vector lengths, which may obscure the geometric interpretation of the resulting expressions. In contrast, our formulation based on cross product tensors provides a more direct geometric correspondence and conveys a clearer physical meaning.

All descriptors in LiTEN are constructed through tensor operations within atomic neighborhoods, enabling the three-body and four-body modules to achieve a linear computational complexity O(N), far surpassing traditional combinatorial approaches. The model demonstrates efficient scalability to large systems and exhibits excellent generalization capabilities across a wide range of chemical environments without relying on predefined atom types. As shown in **Figure 1C-1D**, the foundation model LiTEN-FF, pretrained on NablaDFT and SPICE datasets, powers LiTEN-FF's broad applications, including vacuum-phase conformation optimization, free energy surface (FES) modeling, dihedral scanning, and accurate geometric property prediction. Moreover, the model supports efficient conformer search and is compatible with periodic systems. In summary, LiTEN-FF strikes an optimal balance between accuracy, scalability, and generality, offering a powerful tool for molecular simulations, AI-driven force fields, and large-scale molecular modeling.

**Tests of model accuracy and generalization for quantitative properties**

To systematically evaluate the predictive performance and generalization ability of the LiTEN model across varying molecular scales, we selected two representative standard benchmark datasets: rMD17 and MD22. These datasets cover typical small and large molecular systems, respectively, offering a comprehensive assessment of the model's performance across diverse scenarios in terms of structural complexity, atomic count, and chemical diversity.

The rMD17 dataset consists of multiple representative small organic molecules such as Aspirin, Benzene, and Ethanol. These molecules generally have low molecular weights, relatively rigid conformations, and stable electronic structures and geometries. They are commonly used to test the model's sensitivity and accuracy in capturing subtle atomic interactions within low-dimensional, rigid systems. For these molecules, energy differences mainly arise from minor conformational perturbations, demanding very high accuracy in force prediction. As a result, this dataset serves as an important benchmark for evaluating the model's fine-scale structural modeling capabilities at the small molecular scales. In contrast, the MD22 dataset primarily contains more complex, biologically relevant large molecular systems such as the tripeptide Ac-Ala3-NHMe, carbohydrates like Stachyose, DNA fragments (AT-AT, AT-AT-CG-CG), and supramolecular structures such as the Buckyball Catcher and Double-walled Nanotube (Dw_Nanotube). These systems feature high atom counts, numerous degrees of freedom, and rich non-covalent interactions including hydrogen bonding, π-π stacking, and van der Waals forces. MD22 effectively tests the model's expressive power, scalability, and structural generalization ability in high-dimensional structural spaces. Therefore, its provides a stringent benchmark to determine whether LiTEN can robustly generalize to complex molecular systems.

For comparison, we selected several mainstream advanced DL-driven force field models, which can be roughly divided into two groups. The first group is based on spherical harmonics convolution, such as MACE, NequIP, and BOTNet[40]. These methods explicitly consider rotational equivariance in tensor construction, making them highly suitable for capturing direction-sensitive angles and many-body interactions. The second group relies on on Cartesian coordinate message passing mechanisms, such as PaiNN, Allegro, and VisNet[41], which maintain responsiveness to directional information while generally offering higher efficiency and structural adaptability. Additionally, we included several representative models, namely GemNet[42], ACE[43], ANI, and sGDML[44] as references to cover a broader range of paradigm designs and modeling strategies.

In the experimental setup, we uniformly used the optimal hyperparameter configurations recommended in the literature for each model to ensure fairness in comparison. Except for NequIP and MACE, which used their recommended numbers of layers (6 layers and 2 layers respectively), other models employed a 6-layer network architecture with a hidden dimension size of 256. VisNet specially used a 9-layer network structure to match the settings in its original paper. During the training process, all models were run under the same training-validation-test splits, and standard error metrics (MAE) were used to evaluate the energy (in meV or kcal/mol) and force (in meV/Å or kcal/mol/Å) predictions.

On the rMD17 dataset, LiTEN achieved the best or second-best results in the vast majority of small molecule systems, as shown in **Table 1**. Notably, for well-known molecules such as aspirin, benzene, and ethanol, its force prediction errors were significantly lower than those of other models, demonstrating outstanding performance in high-precision modeling of small perturbations. This result also highlights the efficiency and stability of the model's directional aggregation mechanism and equivariant tensor modeling in capturing subtle conformational changes. On the MD22 dataset, LiTEN showed excellent performance in several structurally complex large molecule systems, as shown in **Table 2**. For example, in systems like Ac-Ala3-NHMe, Stachyose, AT-AT, and Dw_Nanotube, LiTEN significantly outperformed most comparative models in terms of energy and force prediction errors. Especially in scenarios with a large number of atoms, its prediction errors exhibited only negligible fluctuations as the system size increased. This fully demonstrates that LiTEN not only has good scalability when modeling large-scale molecules but also possesses structural robustness and strong generalization ability across different molecular types.

Overall, as a many-body graph neural network with equivariant tensor modeling capabilities, LiTEN achieves near-physical-limit high-precision predictions in small molecule systems and also demonstrates excellent modeling capability and stability in large molecule systems. Compared to existing state-of-the-art models, LiTEN shows significant advantages in terms of energy and force prediction accuracy across multiple test sets and molecular structures. This validates the effectiveness of its proposed directional aggregation, tensorized four-body attention, and implicit structure-aware mechanisms, reflecting strong physical consistency and broad adaptability. In the future, LiTEN is expected to play a pivotal role in complex tasks such as drug design, biomolecular modeling, and new material prediction. It has the potential to emerge as one of the efficient and general-purpose deep learning force field models.

**Table 1.** Mean absolute errors on the rMD17 dataset. Energy (E, meV) and force (F, meV/Å) errors for different models trained on 950 configurations, validated on 50, and evaluated on the held-out test set.

|  |  | LiTEN | MACE | Allegr | VisNet | BOTNet | NequIP | GemNet | AC | ANI | PaiNN |
|---|---|---|---|---|---|---|---|---|---|---|---|
| **Aspirin** | Energy | **1.8** | 2.2 | 2.3 | 1.9 | 2.3 | 2.3 | - | 6.1 | 16.6 | 6.9 |
|  | Force | **6.5** | 6.6 | 7.3 | 6.6 | 8.5 | 8.2 | 9.5 | 17.9 | 40.6 | 16.1 |
| **Azobenzene** | Energy | **0.5** | 1.2 | 1.2 | 0.7 | 0.7 | 0.7 | - | 3.6 | 15.9 | - |
|  | Force | **2.2** | 3.0 | 2.6 | 2.5 | 3.3 | 2.9 | - | 10.9 | 35.4 | - |
| **Benzene** | Energy | **0.03** | 0.4 | 0.3 | **0.03** | **0.03** | 0.04 | - | 0.04 | 3.3 | - |
|  | Force | **0.2** | 0.3 | **0.2** | **0.2** | 0.3 | 0.3 | 0.5 | 0.5 | 10.0 | - |
| **Ethanol** | Energy | **0.3** | 0.4 | 0.4 | **0.3** | 0.4 | 0.4 | - | 1.2 | 2.5 | 2.7 |
|  | Force | 2.4 | **2.1** | **2.1** | 2.3 | 3.2 | 2.8 | 3.6 | 7.3 | 13.4 | 10.0 |
| **Malonaldehyde** | Energy | **0.6** | 0.8 | **0.6** | **0.6** | 0.8 | 0.8 | - | 1.7 | 4.6 | 3.9 |
|  | Force | 4.2 | 4.1 | **3.6** | 3.9 | 5.8 | 5.1 | 6.6 | 11.1 | 24.5 | 13.8 |
| **Naphthalene** | Energy | **0.2** | 0.5 | **0.2** | **0.2** | **0.2** | 0.9 | - | 0.9 | 11.3 | 5.1 |
|  | Force | 1.1 | 1.6 | **0.9** | 1.3 | 1.8 | 1.3 | 1.9 | 5.1 | 29.2 | 3.6 |
| **Paracetamol** | Energy | **1.1** | 1.3 | 1.5 | **1.1** | 1.3 | 1.4 | - | 4.0 | 11.5 | - |
|  | Force | 4.8 | 4.8 | 4.9 | **4.5** | 5.8 | 5.9 | - | 12.7 | 30.4 | - |
| **Salicylic acid** | Energy | **0.7** | 0.9 | 0.9 | **0.7** | 0.8 | **0.7** | - | 1.8 | 9.2 | 4.9 |
|  | Force | 4.1 | 3.1 | **2.9** | 3.4 | 4.3 | 4.0 | 5.3 | 9.3 | 29.7 | 9.1 |
| **Toluene** | Energy | **0.2** | 0.5 | 0.4 | 0.3 | 0.3 | 0.3 | - | 1.1 | 7.7 | 4.2 |
|  | Force | 1.2 | 1.5 | 1.8 | **1.1** | 1.9 | 1.6 | 2.2 | 6.5 | 24.3 | 4.4 |
| **Uracil** | Energy | **0.3** | 0.5 | 0.6 | **0.3** | 0.4 | 0.4 | - | 1.1 | 5.1 | 4.5 |
|  | Force | 2.4 | 2.1 | **1.8** | 2.1 | 3.2 | 3.1 | 3.8 | 6.6 | 21.4 | 6.1 |

**Table 2.** Mean absolute errors on the MD22 dataset. Energy (E, kcal/mol) and force (F, kcal/mol/Å) errors for various models (VisNet, MACE, Allegro, SO3krates[45], TorchMD[46], PaiNN, sGDML), trained on the same number of configurations as sGDML and evaluated on the held-out test set.

| Molecule | atoms |  | LiTEN | VisNet | MACE | Allegro | SO3kra | TorchM | PaiNN | sGDML |
|---|---|---|---|---|---|---|---|---|---|---|
|  |  | Energy | **0.0597** | 0.0636 | 0.0620 | 0.1019 | 0.337 | 0.1121 | 0.1168 | 0.3902 |
| Ac-Ala3-DHA | 42 | Force | **0.0763** | 0.0803 | 0.0876 | 0.1068 | 0.244 | 0.1879 | 0.2302 | 0.7968 |
| | 56 | Energy | 0.0876 | **0.0741** | 0.1317 | 0.1153 | 0.379 | 0.1205 | 0.1151 | 1.3117 |
| | | Force | **0.0554** | 0.0598 | 0.0646 | 0.0732 | 0.242 | 0.1209 | 0.1355 | 0.7474 |
| Stachyose | 87 | Energy | **0.0845** | 0.0915 | 0.1244 | 0.2485 | 0.442 | 0.1393 | 0.1517 | 4.0497 |
| | | Force | **0.0617** | 0.0879 | 0.0876 | 0.0971 | 0.435 | 0.1921 | 0.2329 | 0.6744 |
| AT-AT | 60 | Energy | 0.1263 | **0.0708** | 0.1093 | 0.1428 | 0.178 | 0.1120 | 0.1673 | 0.7235 |
| | | Force | 0.0915 | **0.0812** | 0.0992 | 0.0952 | 0.216 | 0.2036 | 0.2384 | 0.6911 |
| AT-AT-CG-CG | 118 | Energy | 0.1743 | 0.196 | **0.1578** | 0.3933 | 0.345 | 0.2072 | 0.2638 | 1.3885 |
| | | Force | 0.1490 | 0.148 | **0.1153** | 0.1280 | 0.332 | 0.3259 | 0.3696 | 0.7028 |
| Buckyball_Catcher | 148 | Energy | **0.3440** | 0.537 | - | 0.5258 | - | 0.5188 | 0.4563 | 1.1962 |
| | | Force | **0.1143** | 0.201 | - | 0.0887 | - | 0.3318 | 0.4098 | 0.6820 |
| Dw_Nanotube | 370 | Energy | 0.6484 | **0.601** | - | 2.2097 | - | 1.4732 | 1.1124 | 4.0122 |
| | | Force | **0.2139** | 0.292 | - | 0.3428 | - | 1.0031 | 0.9168 | 0.5231 |

## Training Efficiency Evaluation on the Chignolin Dataset

Chignolin is a mini-protein composed of 10 amino acid residues, characterized by a well-defined β-hairpin structure[47]. Due to its compact size and rich conformational diversity, it is widely used as a benchmark dataset for evaluating protein folding, MD simulations, and the performance of machine learning force field models. In this study, we adopted the Chignolin dataset introduced in the VisNet work to evaluate both the training speed and prediction accuracy of various models, with a particular focus on practical efficiency at moderate molecular scales. All models were trained using a batch size of 4.

The evaluation results demonstrate that the LiTEN family of models achieves significantly faster training speeds while maintaining high prediction accuracy[48]. Specifically, LiTEN-128 (with a hidden dimension of 128) achieved an average training time of 0.0227 seconds per batch, the fastest among all the compared models, with a peak memory usage of only 3912 MiB (also the lowest among the tested models). Its force prediction error (Force MAE) was 0.4094 kcal/mol/Å, outperforming all other models except LiTEN-256. The higher-capacity LiTEN-256 model, with a hidden dimension of 256, further improved the prediction accuracy to 0.3641 kcal/mol/Å, with a training time of 0.0312 seconds per batch and memory consumption of 13949 MiB, still substantially lower than those of mainstream models such as Allegro (34977 MiB) and NequIP (22971 MiB).

As shown in **Figure 2**, the LiTEN series demonstrates a well-balanced trade-off among training speed, memory utilization, and prediction accuracy. In particular, LiTEN-128 exhibits strong potential for deployment in high-throughput simulations and resource-limited environments, attributed to its low hardware demands and fast training. Overall, the results on the realistic protein system Chignolin validate LiTEN's advantages in modeling many-body interactions, highlighting its robust generalization and practical applicability.

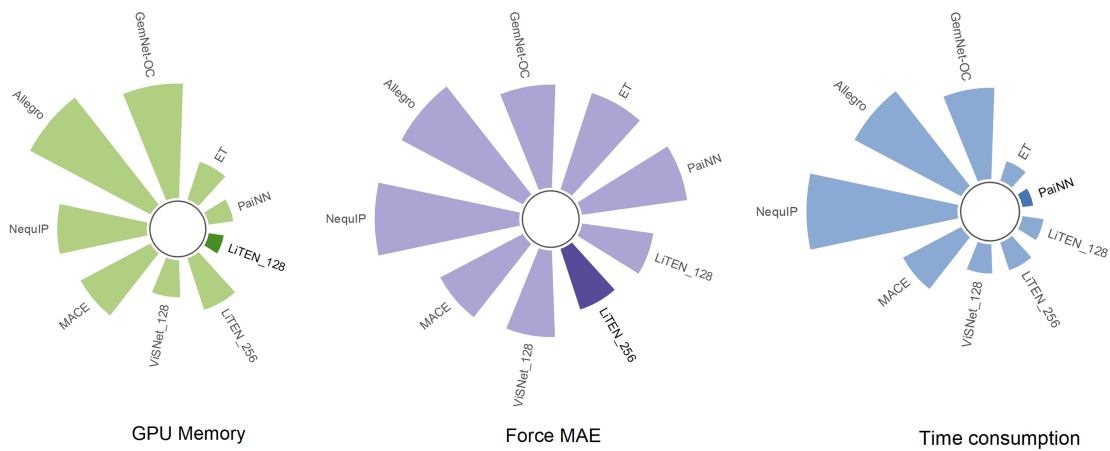

**Figure 2.** Comparison of the force prediction performance and computational efficiency of various models on the Chignolin dataset. Metrics include mean absolute error (MAE, kcal/mol/Å), time consumption (s), and GPU memory usage (MiB). For MAE, lower values signify higher prediction accuracy, while for time and memory consumption, reduced values indicate higher computational efficiency.

## Performance of AIFF foundation model LiTEN-FF

**Hierarchical Two-Stage Training with Large-Scale Quantum Chemical Datasets**

To leverage the full potential of the wealthy large-scale quantum chemical data, we adopted a two-stage hierarchical training strategy. In the initial stage, we performed large-scale pretraining on the nablaDFT dataset, aiming to comprehensively capture the diverse electronic structure patterns prevalent in drug-like molecules. For data partitioning, we reserved geometry optimization trajectories as the test set and selected a subset of samples with significant scaffold diversity and conformational perturbations for hyperparameter tuning. The remaining conformations were used for model training (see **Table S1**), laying a solid foundation for the model's generalization in downstream tasks. Specifically, the model architecture consists of six equivariant convolutional layers, each with 256-dimensional hidden channels, and a cutoff radius of 5 Å, which effectively covers the typical physical range of most intramolecular chemical interactions. The model's equivariance property ensures that the predicted energies and forces adhere to physical symmetries under rotation and translation, thereby improving both its generalization ability and physical consistency. The model was trained using energies in Hartree and forces in Hartree/Å, with a weighted MAE loss. After 100 epochs of training, the model achieved excellent predictive accuracy on the nablaDFT test set, with MAEs of 0.32 milliHatree for energy and 0.20 milliHatree/Å for force, equivalent to approximately 0.20 kcal/mol and 0.126 kcal/mol/Å, respectively. These error levels approach those of high-accuracy quantum chemical methods such as DFT, and notably fall well within the widely accepted threshold (~1 kcal/mol) of what is known as chemical accuracy, demonstrating the potential of LiTEN-FF Nab as a quantum-level surrogate model.

Following pretraining, we conducted fine-tuning on the SPICE dataset, which is characterized by higher precision and presents more significant modeling challenges. Compared to nablaDFT, SPICE offers significantly enhanced molecular diversity and quantum accuracy. We excluded ionic species and non-neutral molecules from the SPICE dataset, ultimately retaining approximately 1.9 million conformations. These were split into a 95:5 train-test ratio, with the training set comprising around 1.8 million conformations, covering molecules with 2 to 110 atoms. In addition to its broader chemical diversity, SPICE includes dimers, solvated biomolecular conformations, and peptides, making it more representative of real-world MD scenarios (as shown in **Table S2**). On the SPICE test set, LiTEN-FF achieved MAEs of 0.57 milliHatree for energy and 0.48 milliHatree/Å for force, corresponding to approximately 0.358 kcal/mol and 0.301 kcal/mol/Å, respectively. Although these errors are slightly higher than those observed on the nablaDFT test set, they still fall within the DFT-level error range, underscoring the model's ability to maintain high predictive accuracy even under distributional shifts. It is noteworthy that all evaluations in this study were conducted using a molecule-level error metric, i.e., MAE computed over the entire molecular energy and forces. This assessment strategy better reflects the model's real-world performance in applications such as drug-like molecule modeling, conformer optimization, and MD simulations. During the SPICE finetuning phase, we particularly focused on evaluating the model's ability to adapt to variations in molecular scaffolds, conformational perturbations, and changes in electronic environments. Through this fine-grained optimization on a high-accuracy dataset, we further enhanced the model's predictive accuracy and robustness in real drug discovery tasks. This approach effectively transitions the model from broad-coverage general learning to high-precision specialized modeling.

**Exploration of Molecular Dynamics Simulations in Vacuum Systems with LiTEN-FF**

In molecular simulations, efficiently and accurately obtaining molecular conformations, geometric parameters, and torsional energy profiles is a fundamental capability that underpins core tasks in drug design and materials research[49]. Conformer optimization is essential for identifying stable molecular structures and serves as the foundation for downstream applications such as docking and property prediction. This is particularly critical in high-throughput drug screening scenarios, where both speed and accuracy are imperative[50]. During MD simulations, the monitering of fluctuations in bond lengths and angles offers valuable insight into a molecule's thermodynamic stability and structural flexibility, making it a key indicator for evaluating the physical fidelity of force fields. Furthermore, torsion angle scanning enables detailed a detailed characterization of intramolecular energy variations, playing an indispensable role in conformational sampling, potential energy surface construction, and force field parameterization. Therefore, focusing on these three critical tasks, we systematically evaluated the applicability of the machine learning-based LiTEN-FF model under vacuum conditions. The assessment encompasses two primary variants of the LiTEN-FF model, namely LiTEN-FF Nab and LiTEN-FF SPICE, evaluated across three core tasks: conformer optimization, geometric property analysis, and torsional energy profile prediction.

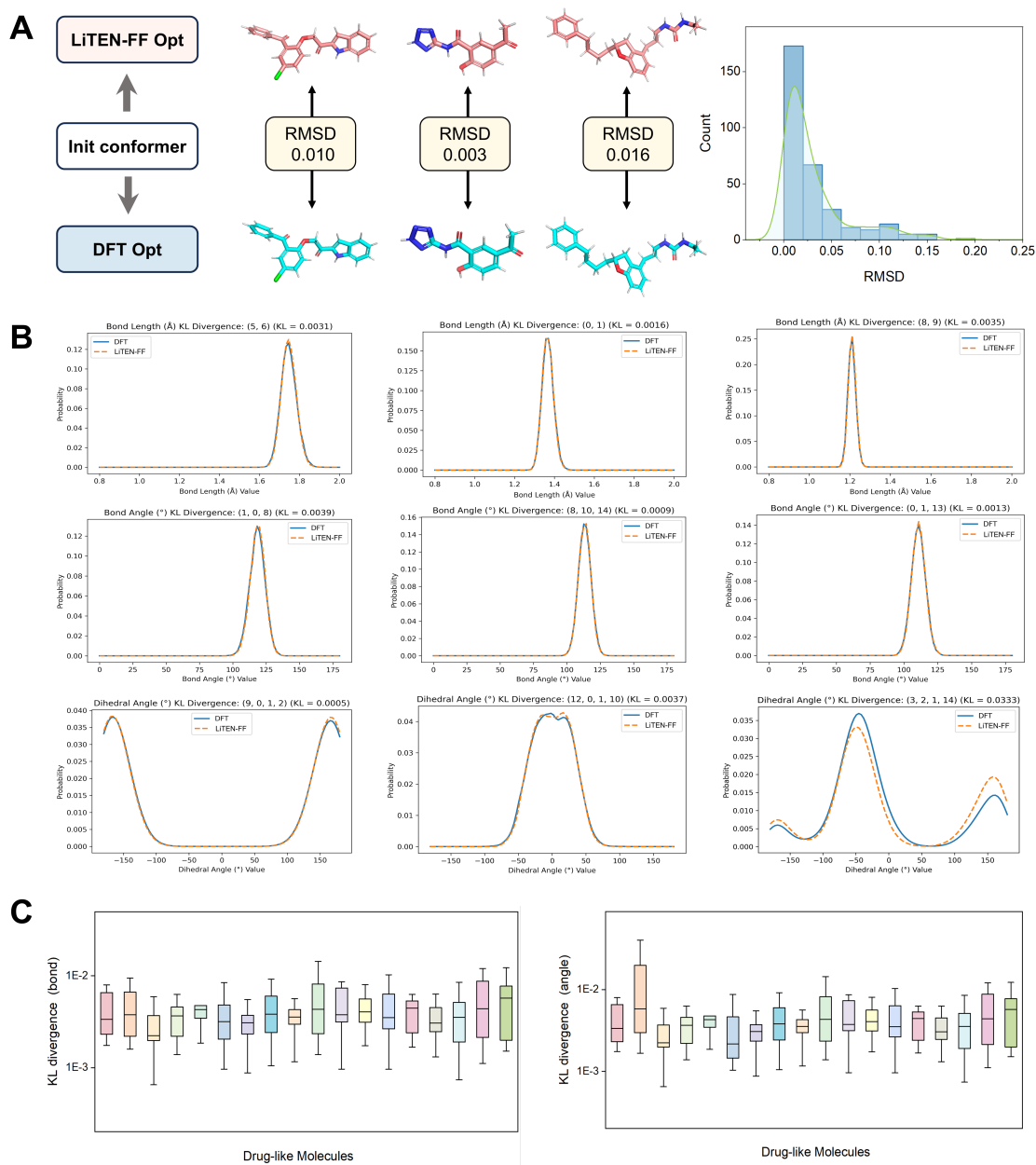

**Figure 3.** Applications and evaluations in vacuum systems. (A) Conformational optimization of 300 molecules starting from their initial geometries, performed using both DFT (ωB97X-D/def2-SVP) and LiTEN-FF Nab. The panel shows examples of RMSD between the final optimized structures and the overall RMSD distribution. (B) Geometric analysis of 18 selected molecules, focusing on bond lengths, bond angles, and dihedral angles. Distributions obtained from both DFT and LiTEN-FF are presented for representative molecules, demonstrating highly consistent results between the two methods. (C) KL divergence between the DFT and LiTEN-FF distributions for all bond lengths and bond angles across the 18 molecules, highlighting the close agreement between the two computational approaches.

**Conformation Optimization Using LiTEN-FF Nab.** To systematically assess the applicability of LiTEN-FF Nab in the conformer optimization of drug molecules, we conducted an in-depth performance analysis across multiple dimensions. A test set of 300 molecules, with atom counts ranging from 8 to 62, was selected. These molecules were excluded from the training set, thereby ensuring both the validity and representativeness of the evaluation[51]. This size range covers the majority of commonly encountered drug-like compounds, lending the study substantial practical relevance.

All initial conformers were uniformly generated using RDKit[52], which eliminates biases arising from different starting geometries and enables a fair comparison among various methods. The optimization process was carried out through the ASE interface with a uniform convergence threshold of 0.05 eV/Å for the maximum force, a stringent criterion that ensures high-precision conformer optimization. In terms of accuracy, the conformers optimized by LiTEN-FF Nab demonstrated excellent agreement with the reference structures optimized by DFT ($\omega$B97X-D/def2-SVP). The mean RMSD between the two sets of structures was 0.05 Å, indicating the model's ability to accurately reproduce DFT-optimized geometries. More specifically, as shown in **Figure 3A**, 31.2% of the molecules had RMSD values below 0.01 Å, indicating near-exact reproduction of DFT geometries. Furthermore, 79.7% of the molecules had RMSD values below 0.05 Å, demonstrating very low structural deviation for the majority of the test set. Additionally, 89.8% had RMSD values below 0.1 Å, further validating the overall robustness and reliability of the optimized results.

The energy profiles of the optimization trajectories generated by LiTEN-FF Nab closely matched those obtained from DFT, implying that the model not only accurately locates energy minima but also captures the energy landscape along the optimization path. This capability is essential for tasks such as conformer searching and MD simulations. Notably, LiTEN-FF Nab demonstrated over a 1000-fold speedup compared to high-accuracy DFT methods. This substantial boost in computational efficiency could greatly reduce the drug discovery timeline and provides a strong technical foundation for high-throughput screening and virtual screening pipelines. While the current evaluation is limited to molecules composed of elements included in the training set, future work will aim to extend the model to systems containing metal atoms and other complex functional groups. Incorporating solvent effects and temperature dependence into the optimization process also represents a promising direction for exploring the model's stability under more realistic conditions. Furthermore, integrating LiTEN-FF Nab with MD simulations to evaluate its performance in capturing dynamic conformational changes over extended timescales could advance its applications in drug design and materials science.

**Conformer Geometry Calculations.** To further assess LiTEN-FF's ability to model dynamic geometric behavior, we conducted MD simulations under vacuum conditions on 18 organic molecules not present in the training data. These molecules represent a range of sizes (≤ 25 atoms) and contain typical organic elements such as C, H, O, N, F, Cl, Br, and S. The simulations, lasting between 50 to 150 ps per molecule, were designed to test the model's fidelity in reproducing time-evolving molecular geometries and internal coordinate distributions.

A comparative statistical analysis was conducted between the LiTEN-FF-predicted geometry distributions and those derived from the DFT-based MD simulations. Specifically, we computed Kullback-Leibler (KL) divergence metrics for bond lengths and bond angles across all molecules. As shown in **Figures 3B-3C**, the KL divergence remained below 0.01 in nearly all cases, indicating that LiTEN-FF closely reproduces the geometric ensembles generated by DFT. Bond length distributions showed tight peaks around equilibrium values, and angular distributions captured both symmetric and asymmetric features of molecular flexibility. Furthermore, LiTEN-FF was able to maintain overall molecular stability and structural integrity across simulation timescales, without introducing artificial bond distortions or unphysical behaviors, an issue sometimes observed in less robust ML potentials. We also present molecule-specific statistical summaries **(Table S2)** and visualizations **(Figures S1-S2)** of the geometric property distributions, including variance, skewness, and kurtosis. These analyses reveal consistent trends: smaller molecules typically exhibit sharper and more narrowly clustered distributions, while slightly larger or more flexible molecules show broader but still DFT-consistent behavior. This suggests that LiTEN-FF not only accurately captures equilibrium geometries but also reflects the intrinsic thermal fluctuations in molecular conformations, making it suitable for dynamic simulations that require physical realism over extended timescales.

**Torsion Angle Scanning: TorsionNet206 Dataset.** Performance comparison on the TorsionNet206 benchmark reveals key insights into the ability of neural network potential models to accurately reproduce torsional energy profiles, a task that plays a vital role in drug design. Traditional density functional theory methods, such as ωB97M D3BJ combined with the def2 TZVPPD basis set, remain the most accurate, achieving a MAE of 0.15 kcal per mol and serving as the reference standard. Nevertheless, recent neural network models such as Egret One and the MACE OFF series have shown excellent performance, with MAEs ranging from 0.19 to 0.24 kcal per mol and correlation coefficients exceeding 0.98 for both R squared and Spearman metrics. These results greatly reduce the accuracy gap with DFT while delivering significant improvements in computational efficiency.

Notably, as shown in **Table 3**, our LiTEN-FF SPICE model achieved an MAE of 0.19 kcal/mol on this task, outperforming not only semi empirical approaches such as GFN2 xTB but also traditional DFT methods like B3LYP-D3BJ/6-31G(d). This highlights LiTEN-FF SPICE's ability to effectively capture torsional energy landscapes shaped by complex electronic and steric interactions. The pretrained LiTEN-FF Nab model, while slightly less accurate with an MAE of 0.54 kcal/mol, still outperforms many earlier neural models. This highlights the critical role of training data quality, as the accuracy of the model is closely tied to that of its training targets. The substantial performance improvement observed when LiTEN-FF Nab is fine-tuned to LiTEN-FF SPICE clearly illustrates the importance of domain-specific fine-tuning.

Importantly, the consistently high correlation metrics across different LiTEN-FF variants demonstrate that the models not only provide accurate absolute energy predictions but also preserve the correct ranking of torsional conformers. This is essential for downstream applications such as conformer selection and force field parameterization.

In conclusion, LiTEN-FF SPICE delivers DFT-level accuracy in high-throughput torsion scan tasks while drastically reducing computational costs. Compared to models like Egret-1 and MACE_OFF, LiTEN-FF also offers a significant speed advantage since it does not rely on spherical harmonic tensor computations. These results position LiTEN-FF SPICE as a practical and efficient alternative for torsional energy modeling in real-world molecular design workflows.

Table 3. Evaluation on Torsion206 Dataset Compared with Baseline Models (Unit: kcal/mol)

| Method | Theory | MAE | $R^2$ | Spearman |
|---|---|---|---|---|
| ωB97M-D3BJ/def2-TZVPPD | DFT | 0.15 | 0.99 | 0.98 |
| B97-3c | DFT | 0.35 | 0.98 | 0.97 |
| r²SCAN-3c | DFT | 0.42 | 0.97 | 0.97 |
| B3LYP-D3BJ/6-31G(d) | DFT | 0.57 | 0.95 | 0.94 |
| GFN2-xTB | SE | 0.73 | 0.85 | 0.85 |
| AIMNet2 | NNP | 0.39 | 0.95 | 0.94 |
| MACE-MP-0b2-L[53] | NNP | 1.15 | 0.74 | 0.75 |
| Orb-v3[54] | NNP | 0.97 | 0.83 | 0.83 |
| OMat24 eqV2-L[55] | NNP | 1.48 | 0.77 | 0.81 |
| Egret-1[56] | NNP | 0.20 | 0.99 | 0.98 |
| Egret-1e | NNP | 0.22 | 0.99 | 0.98 |
| Egret-1t | NNP | 0.23 | 0.99 | 0.98 |
| MACE_OFF (23_medium) | NNP | 0.24 | 0.99 | 0.99 |
| MACE_OFF (23_large) | NNP | 0.21 | 0.99 | 0.99 |
| MACE_OFF (24_medium) | NNP | 0.27 | 0.98 | 0.99 |
| **LiTEN-FF Nab** | **NNP** | **0.54** | **0.93** | **0.96** |
| **LiTEN-FF SPICE** | **NNP** | **0.19** | **0.99** | **0.99** |

## LiTEN-FF MD Speed and Periodic Water Box Simulation

To comprehensively evaluate the computational efficiency of LiTEN-FF, we conducted a systematic benchmark study against several widely used neural force field models, including MACE-OFF(L), MACE-OFF(M), DPA2, and AIMNET2, across molecule containing 100 to 1500 atoms. As shown in **Figure 4A**, LiTEN-FF consistently demonstrates the lowest computational overhead across all system sizes. For smaller systems containing up to 300 atoms, it completes 10,000 MD steps in less than 200 seconds, outperforming Mace Large, which starts with a runtime of approximately 100 seconds but rises sharply to over 500 seconds as the system size exceeds 300 atoms. Similarly, Mace Medium exceeds 400 seconds beyond this scale. For larger systems, such as those with 1200 atoms, LiTEN-FF maintains a runtime below 400 seconds, achieving up to a ten-fold speedup compared to Mace Large, which exceeds 3500 seconds for the same system size.

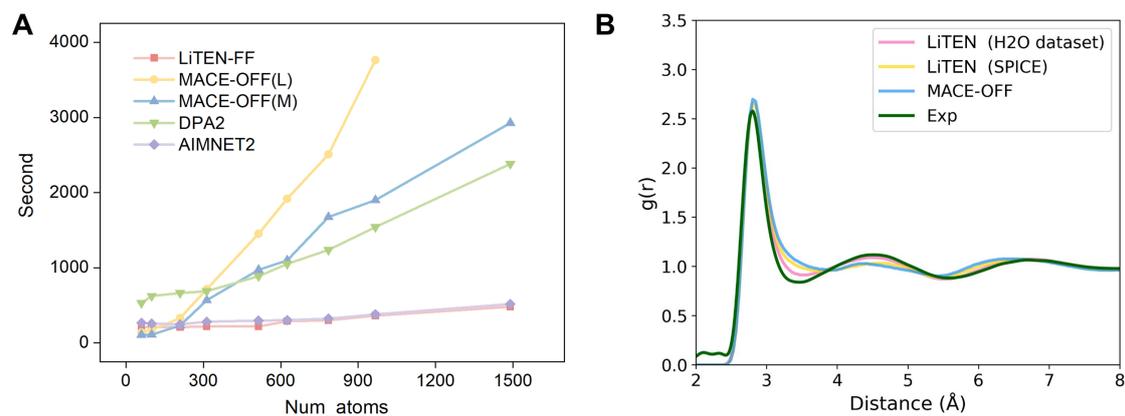

**Figure 4.** (A) Runtime comparison of LiTEN-FF and baseline models for 10,000 MD steps across systems with varying atom counts. (B) Performance of LiTEN-FF in a periodic water box, as shown by the radial distribution function between oxygen atoms (O–O).

This performance gap becomes increasingly evident as the system size expands, highlighting the scalability advantages of LiTEN-FF. While other models tend to experience a rapid increase in computational cost as molecular size grows, LiTEN-FF maintains efficient inference through its TQA mechanism, which compactly encodes higher-order geometric correlations with linear computational complexity. This architectural efficiency allows LiTEN-FF to scale gracefully, offering substantially faster runtimes without sacrificing predictive accuracy.

Beyond speed, LiTEN-FF achieves efficiency without compromising physical fidelity. To validate this, we evaluated LiTEN-FF's performance in periodic systems using a cubic water box consisting of 348 water molecules. Given water's critical role as both a fundamental molecular system and a benchmark for evaluating biomolecular force fields, accurate modeling under periodic boundary conditions is essential. To this end, we explored finetuning LiTEN-FF on the high-fidelity dataset from Chen et al.[57], which captures accurate thermodynamic behavior across diverse molecular configurations. We then conducted 10 ns MD simulations to calculate the radial distribution function (RDF) of oxygen-oxygen distances **(Figure 4B)**. The LiTEN-FF model finetuned on the ab initio water dataset yielded RDFs that closely aligned with experimental measurements[58], accurately capturing both the first RDF peak and the subsequent hydration shells. In contrast, models trained solely on the SPICE dataset, such as LiTEN-FF (SPICE) and MACE-OFF, demonstrated nearly identical deviations, including an overestimation of the first hydration valley (3-4 Å) and an underestimation of the first RDF peak (around 4.5 Å). These systematic discrepancies are primarily attributed to the limitations of the SPICE dataset, particularly its insufficient sampling of the water configurational space and limited thermodynamic accuracy.

Taken together, these results confirm that LiTEN-FF offers a compelling balance between computational speed and physical fidelity, especially in large-scale and periodic systems. Its linear-scaling architecture not only reduces simulation cost but also enables high-throughput exploration of complex molecular environments, making it a strong candidate for next-generation machine learning force fields in both vacuum and condensed-phase simulations.

## Assessment of LiTEN-FF for Free Energy Surface Modeling in Vacuum and Periodic Solvent Conditions

The free energy surface (FES) is a fundamental concept for characterizing the conformational dynamics of biomolecules. It quantifies the relative stability of different conformations and maps the kinetic pathways and energy barriers between them. Accurate reconstruction of the FES is essential for elucidating mechanisms such as protein folding, enzymatic reactions, and ligand binding. However, achieving both high accuracy and computational efficiency in FES calculations remains a major challenge in molecular simulations and a critical benchmark for assessing force fields and machine learning potentials. In this study, the performance of the LiTEN-FF model is evaluated in the context of FES prediction, with a focus on its sampling efficiency and energy estimation accuracy.

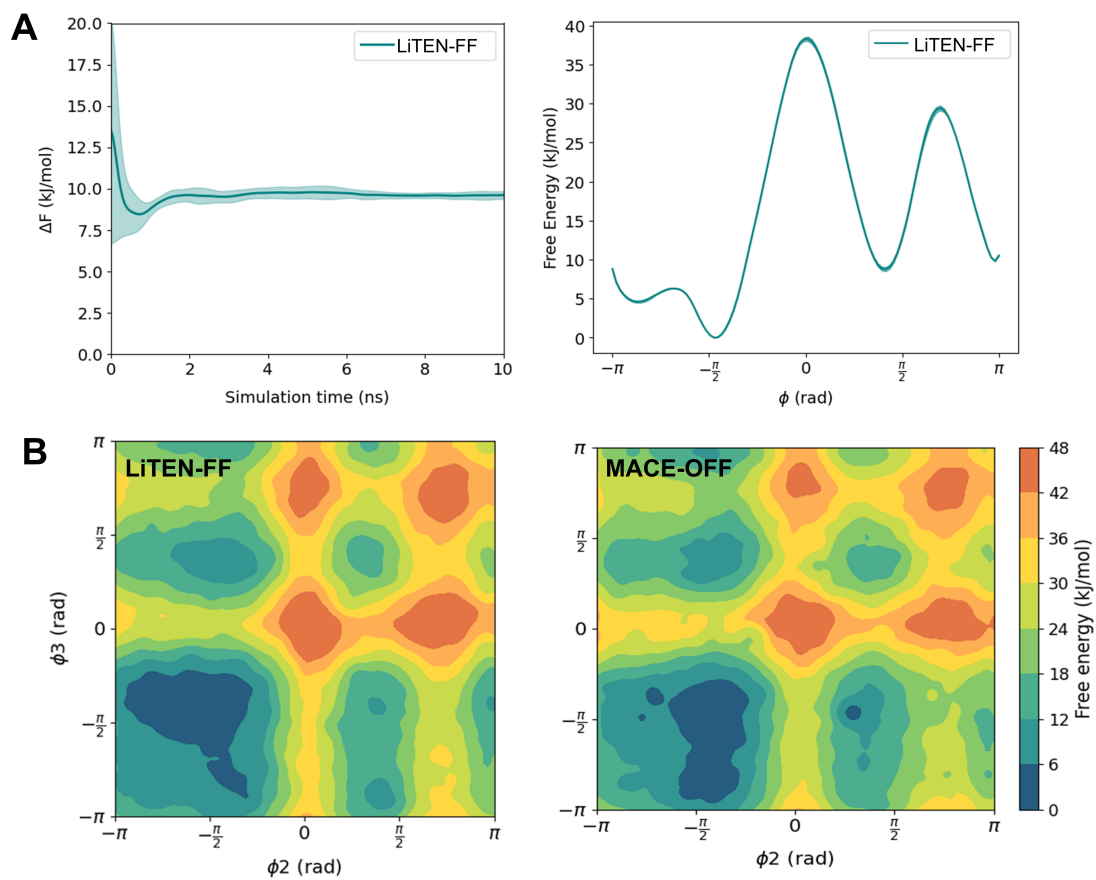

**Figure 5**. (A) Convergence of free energy differences over time and the corresponding free energy surface of alanine dipeptide. The solid line denotes the mean value, while the dashed lines represent the standard deviations from three independent simulations. (B) Free energy surface of alanine tetrapeptide illustrating key metastable conformations. The three-dimensional structures shown at the bottom left, bottom right, top left, and top right correspond to these distinct metastable states, respectively.

**Free Energy Surface Construction and Performance for Alanine Dipeptide (ala2).** The alanine dipeptide (ala2) has long been regarded as a canonical system for testing biomolecular simulation methods due to its well-characterized conformational states. Utilizing the LiTEN-FF model, we performed 10 ns enhanced sampling simulations to characterize the backbone dihedral angles $\varphi$ and $\psi$, as shown in **Figure 5A**. The implementation of the on-the-fly probability enhanced sampling (OPES) technique accelerated the exploration of rarely visited conformations, enabling the rapid convergence of free energy surface[59]. Across three independent parallel trajectories, the free energy difference between key metastable states was approximately 9.62 kJ/mol, with a statistical uncertainty significantly lower than $0.5\,k_b T$, underscoring the model's excellent reproducibility and statistical robustness. Crucially, the free energy profile predicted by LiTEN-FF aligns nearly perfectly with high-level DFT calculations, therefore validating the model's outstanding accuracy. This achievement demonstrates LiTEN-FF's ability to combine theoretical rigor with computational efficiency, providing a reliable foundation for simulating more complex molecular systems.

**Free Energy Surface Simulation and Model Validation for Alanine Tetrapeptide (ala4) in Aqueous Solution.** Building on its proven performance in vacuum systems, LiTEN-FF was further applied to the more challenging system of alanine tetrapeptide (ala4) in aqueous solution. This peptide exhibits extensive conformational flexibility governed primarily by three backbone dihedral angles $\varphi_1, \varphi_2, \text{and } \varphi_3$[59]. We conducted 20 ns OPES simulations using these dihedrals as collective variables, and our focus was on the slower-relaxing $\varphi_2 - \varphi_3$ subspace to reveal the dominant conformational basins. The free energy landscapes generated from LiTEN-FF (fine-tuned on the SPICE dataset) and the MACE-OFF model show remarkable agreement, consistently identifying four metastable states, as shown in **Figure 5B**. The basin defined by $\varphi_2 < 0$ and $\varphi_3 < 0$ corresponds to the lowest free energy minimum, and it features a structure akin to an antiparallel β-sheet, thereby representing the peptide's most stable conformation. Notably, LiTEN-FF predicts a slightly broader basin compared to MACE-OFF, suggesting a more nuanced sampling of conformational heterogeneity. Furthermore, LiTEN-FF exhibits superior computational efficiency relative to MACE-OFF, highlighting its practical advantages for large-scale biomolecular simulations. These results collectively confirm LiTEN-FF's robustness and predictive power in accurately reproducing complex biomolecular free energy landscapes within realistic solvent environments.

## Conformer search with LiTEN-FF

Generating multiple conformers of drug-like molecules is essential in drug design, as it enables accurate modeling of ligand-target interactions by capturing the molecular flexibility critical for binding and biological activity. After demonstrating the reliability of LiTEN-FF in molecular simulations, we applied it to the task of conformer generation. Specifically, we performed multiple rounds of high-temperature and annealing simulations on an entire batch of diverse molecules to explore the conformational space. After each annealing cycle, we optimized the resulting structures using the FIRE algorithm[60] to obtain well-relaxed conformers. The initial coordinates for a new simulation round were taken from the last frame of the previous high-temperature simulation.

To determine whether the generation process had converged, we introduced a convergence criterion on RMSD analysis. For each newly generated conformer, we computed the RMSD against all previously generated conformers of the same molecule. If, over a certain number of consecutive rounds, every newly generated conformer could be matched to an existing conformer within a predefined RMSD threshold, we considered the generation to have converged and terminated further simulations. This ensured that no substantially new conformations were being sampled, thereby indicating sufficient coverage of the conformational space. After the simulations were completed, we removed redundant conformers based on the predefined RMSD threshold.

To demonstrate the efficacy of of our method in thoroughly exploring the conformational space, we performed t-SNE dimensionality reduction on the conformers generated for a benchmark set of 50 molecules and visualized the results using UMAP plots **(Figure 6A)**. These conformations not only exhibit structural diversity but also span across distinct energy basins, indicating broad coverage of the underlying energy landscape. The widespread distribution of points in the UMAP plot highlights the method's strength in capturing both high- and low-energy conformers, reflecting its ability to overcome energy barriers and explore multiple metastable states. We evaluated the efficiency of our conformer generation workflow using 1, 10, 30, 50, and 100 drug-like molecules with distinct structures. Each test involved 250 simulation rounds, with each round consisting of 100 steps of high-temperature dynamics followed by 100 steps of annealing. Additionally, the RMSD threshold for the convergence was detected in five consecutive rounds. Notably, the inference speed of LiTEN-FF does not scale linearly with the number of atomic coordinates, making our approach particularly well-suited for large-batch conformer generation. As shown in **Figure 6B**, increasing the batch size to 100 accelerates the conformer generation process by approximately 10-fold compared to generating conformers one-by-one. All benchmarks were performed on an NVIDIA GeForce RTX 4090 GPU with 24 GB of memory.

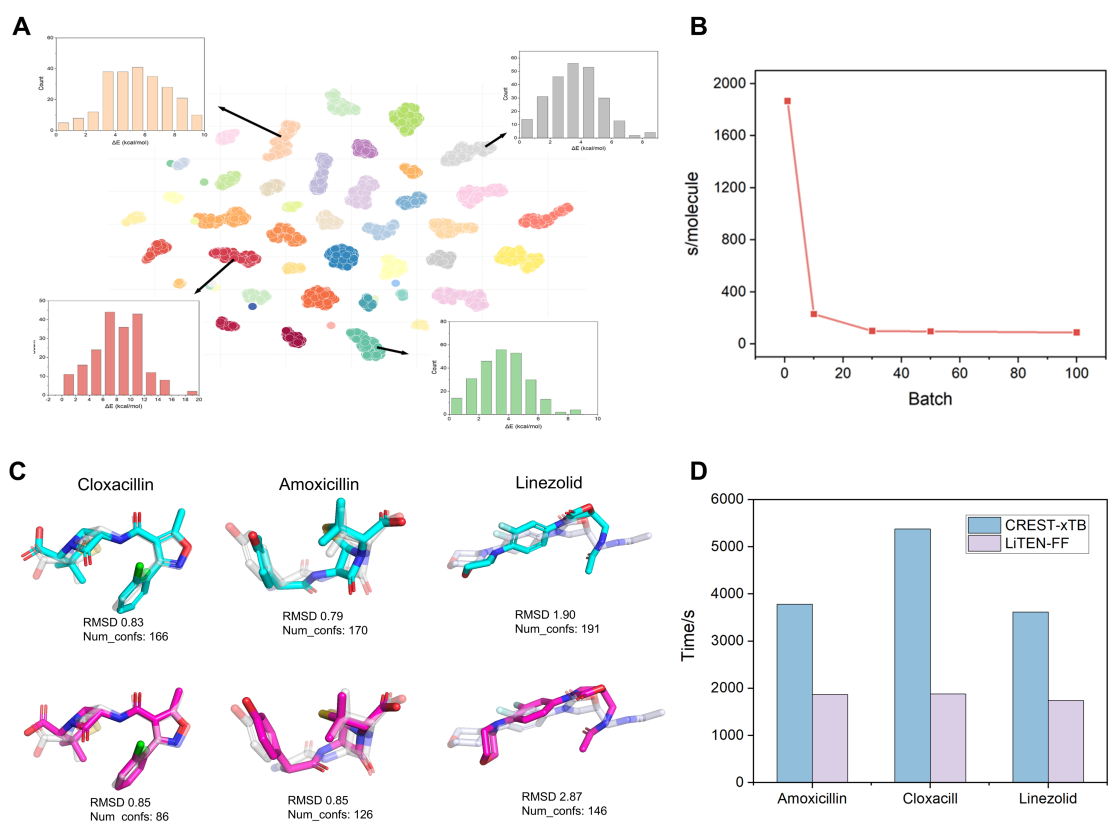

**Figure 6**. (A) Conformational distributions obtained from a set of molecules within a single batch during conformer search. (B) Time required to complete conformer search per molecule as the batch size increases. (C) Comparison of the lowest-energy conformers identified by LiTEN-FF and CREST-xTB, along with the RMSD values relative to the initial structures. (D) Conformer search time for three representative molecules using LiTEN-FF and CREST-xTB, respectively.

For single-molecule tasks, we evaluated three representative drug molecules:

Amoxicillin, Cloxacillin, and Linezolid, and compared our results with those obtained using the CREST level of theory. As illustrated in **Figure 6C**, our method generates a greater number of diverse conformers than CREST, capturing a broader range of structure variations. Additionally, the conformers produced by our approach exhibit more detailed sampling, as reflected by lower RMSD values and better configurational convergence. In terms of computational efficiency, our method is markedly faster, requiring less than half the runtime of CREST across all three molecules **(Figure 6D)**.

In summary, our generative method is capable of rapidly generating a large number of reliable conformations for a wide range of drug-like molecules, significantly advancing the efficiency and effectiveness of conformational sampling in drug discovery.

## Conclusion

The development of LiTEN-FF marks a significant milestone in the field of molecular simulation. As an AI-based foundational potential model specifically designed for biomolecular systems, LiTEN-FF is built upon the equivariant tensorized neural network architecture of LITEN. It employs the TQA mechanism in Cartesian coordinates to efficiently model many body interactions, thereby avoiding the computational overhead associated with spherical harmonic expansions. This architecture has demonstrated SOTA accuracy across multiple from-scratch benchmarks, including rMD17, MD22, and Chignolin, validating its effectiveness. LiTEN-FF was pretrained on the large-scale nablaDFT dataset, which encompasses a variety of drug-relevant elements, endowing it with strong generalization capabilities. Further fine-tuning on the high-precision SPICE dataset enhances its adaptability to solvated environments. By supporting downstream tasks such as conformer search, geometry optimization, and free energy surface construction, LiTEN-FF achieves up to a tenfold speedup compared to MACE on systems with thousands of atoms, significantly improving the feasibility of large-scale molecular simulations and providing a solid foundation for end-to-end drug discovery workflows. The impact of LiTEN-FF is expected to extend far beyond its initial applications, potentially driving a paradigm shift in computational chemistry and drug design. In future versions, we plan to incorporate techniques such as knowledge distillation to further compress the model size, accelerate inference speed, and expand application scenarios for more efficient deployment. Coupled with ongoing advancements in computational power and deeper integration with protein structure prediction and molecular generation models, LiTEN-FF is poised to become a core engine for AI-driven molecular simulation, further pushing the boundaries of drug design and materials science.

## Methods

### Invariance of Energy and Equivariance of Forces

In the fields of MD and geometric deep learning, it is crucial to ensure that the physical quantities predicted by models satisfy fundamental symmetry requirements[61]. This section provides rigorous mathematical proofs demonstrating the fundamental symmetry properties of the key physical quantities in our framework.

The scalar energy function exhibits rotational invariance, satisfying:

$$E(Ru) = E(u). \tag{8}$$

for any rotation matrix $R$. This invariance stems from two crucial properties:

Vector norm preservation: $|Ru_i| = |u_i|$, Angle cosine preservation:

$$\cos \theta'_{jik} = \frac{Ru_{ji} \cdot Ru_{ki}}{|Ru_{ji}||Ru_{ki}|} = \cos \theta_{jik} \tag{9}$$

The cross products term maintains rotational invariance:

$$(u_j \times u_{jk}) \cdot (u_{jk} \times u_k) \tag{10}$$

Through the transformation property $(Ra) \times (Rb) = R(a \times b)$ and orthogonality $R^T R = I$, we establish:

$$[(Ru_j) \times (Ru_{jk})] \cdot [(Ru_{jk}) \times (Ru_k)] = (u_j \times u_{jk}) \cdot (u_{jk} \times u_k) \tag{11}$$

The cubic norm expression inherits rotational invariance from its constituent invariant terms:

$$|u_i|^3 = \left(\sum \cos \theta_{jik}\right) \cdot |u_i| \tag{12}$$

Force equivariance follows from energy invariance through gradient transformation:

$$F(Ru) = -\nabla_{Ru} E(Ru) = -R\nabla_u E(u) = RF(u) \tag{13}$$

where we use the covariant gradient transformation $\nabla_{Ru} = R\nabla_u$. These mathematical guarantees ensure physical consistency under rotational transformations, which can be empirically verified by applying random rotations to input coordinates and testing the invariance/equivariance conditions. This theoretical foundation supports the development of reliable geometric learning models for molecular simulations.

**Tensorized Quadrangle Attention**

In the **Model Architecture** section, we proposed an efficient and physically grounded mechanism to capture dihedral interactions in molecular graphs, termed Tensorized Quadrangle Attention (TQA). This method achieves linear complexity while maintaining high expressiveness by implicitly modeling four-body geometric relationships during edge feature updates. The TQA mechanism is formulated as:

$$E_{ij}^l = \left[(\vec{u_i} \times \vec{u_{ij}}) \cdot (\vec{u_{ij}} \times \vec{u_j})\right] \odot \text{SiLU}(W_e E_{ij}^{l-1} + b_e) \tag{14}$$

where $\vec{u_i}$ and $\vec{u_j}$ are the directional vector embeddings of nodes $i$ and $j$, and $\vec{u_{ij}}$ is the normalized direction vector from node $i$ to node $j$. The term $(\vec{u_i} \times \vec{u_{ij}}) \cdot (\vec{u_{ij}} \times \vec{u_j})$ captures a torsion-like relationship, analogous to a dihedral angle, but implemented in a differentiable and vectorized form. This quantity modulates the updated edge features through a Hadamard (element-wise) product with a nonlinear transformation of the previous edge state.

This formulation allows us to encode localized quadrangle interactions across edges, implicitly modeling the torsional coupling between bonded atoms without the need for an explicit enumeration of four-body terms. Since each message-passing step already includes information from both node directions and edge vectors, the effective receptive field is enlarged. With a cutoff radius of 5 Å for edge construction, this tensorized attention spans roughly 10 Å in a single layer, enabling the network to capture mid- to long-range structural correlations efficiently.

To incorporate node-level information and further enrich the representation, we apply a learned attention mechanism. Specifically, attention scores are computed as:

$$\text{Attention}_{ij}^l = \text{SiLU}(h_i^l + E_{ij}^l + h_j^l) \odot \text{Alpha} \tag{15}$$

where $h_i^l$ and $h_j^l$ are the scalar features of atoms $i$ and $j$, and $Alpha$ is a trainable parameter vector that modulates attention weights. This attention is then gated by a continuous cutoff function based on the interatomic distance:

$$\text{Attention}_{ij}^l = \text{Attention}_{ij}^l \odot \text{Cutoff}(Distance_{ij}^l) \tag{16}$$

This hybrid design enables adaptive feature weighting that respects both learned chemical relevance and geometric locality. The updated node and vector features are then computed as:

$$h_j^l = h_j^l \odot E_{ij}^l \odot \text{Attention}_{ij}^l, \ V_j^l = V_j^l \odot h_j^l + V_{ij}^l \odot h_j^l \tag{17}$$

Finally, both scalar and vector features are aggregated across neighbors using scatter-based summation and residual connections:

$$h^l = \text{scatter}_{\text{sum}}(h_j^l) + h^{l-1}, \ V^l = \text{scatter}_{\text{sum}}(V_j^l) + V^{l-1} \tag{18}$$

This construction results in a fully vectorized, geometry-aware message passing scheme that models complex conformational dependencies in a scalable and differentiable manner. By capturing implicit four-body interactions through local vector operations, Tensorized Quadrangle Attention enhances the model's capacity to represent torsional energy landscapes and anisotropic molecular features, all while maintaining efficiency suitable for large-scale systems.

## Scalar-Vector Fusion

This component serves to further integrate the vector and scalar node representations derived from the Message Passing and TensorCrossFusion modules. The fusion mechanism is designed to leverage both three-body and higher-order geometric interaction cues, whose theoretical underpinnings have been rigorously justified in the **Model** section. For each node $i$, the vector feature $V_i^l \in R^{C \times 3}$ and scalar feature be

$h_i^l \in R^C$ are processed to capture complex geometric correlations and reinforce the bidirectional exchange of information between vectorial and scalar modalities. To this end, the vector representation is first projected into two separate subspaces:

$$\left[V_i^{l(1)}, V_i^{l(2)}\right] = Linear_{vec}(V_i^l), \quad V_i^{l(1)}, V_i^{l(2)} \in R^{C \times 3} \tag{19}$$

These subspace projections are used to construct two geometric interaction terms: a vector trilinear term, which captures directional alignment (akin to three-body angular interaction), and a vector multi-body term, which encodes magnitude-based modulation (analogous to multi-body geometric scaling). These terms are formally defined as:

$$VecTri_i = \sum_{k=1}^{3} V_{i,:,k}^{l(1)} \odot V_{i,:,k}^{l(2)} \in R^C \tag{20}$$

$$VecMuti_i = \left(\sqrt{\sum_{k=1}^{3} \left(V_{i,:,k}^{l(2)}\right)^2 + \varepsilon}\right)^3 \in R^C, \quad \varepsilon = 10^{-8} \tag{21}$$

These vector-derived signals are used to modulate scalar features. The scalar representation is projected as:

$$\left[h_i^{l(1)}, h_i^{l(2)}, h_i^{l(3)}\right] = Linear_{sca}(h_i^l), \quad h_i^l \in R^C \tag{22}$$

Then, we compute the update residues:

Scalar update residue:

$$h_i^{l+1} = (VecTri_i + VecMuti_i) \odot h_i^{l(1)} + h_i^{l(2)} \tag{23}$$

Vector update residue:

$$V_i^l = V_i^{l(1)} \cdot h_i^{l(3)} \tag{24}$$

where the scalar $h_i^{l(3)} \in R^C$ is broadcast across the spatial dimension of $V_i^{l(1)}$.

These update residues are added back to the original features in subsequent stages, enabling tightly coupled scalar-vector representations that are sensitive to both directional geometry and feature semantics.

## Dataset Construction

NablaDFT is a high-quality quantum chemical dataset specifically curated for pharmaceutical modeling. After integrating multiple of its subsets, we obtained approximately 1.9 million molecules and 16 million conformations. Each conformation is annotated with key quantum chemical properties, including total energy, atomic forces, Hamiltonian matrices, and overlap matrices, calculated at the ωB97X-D/def2-SVP level of theory. The dataset covers a wide range of representative drug-like molecules with atom counts ranging from 8 to 62, and includes eight common pharmaceutical elements: H, C, N, O, S, Cl, F, and Br.

SPICE is a recently released, high-fidelity quantum chemical dataset designed to benchmark the generalization ability of AI-based force fields for small-molecule modeling. It includes over 110,000 small molecules and ~2 million conformations, covering 17 elements: H, Li, B, C, N, O, F, Na, Mg, Si, P, S, Cl, K, Ca, Br, and I. Each conformation is annotated with high-accuracy quantum properties (energy, atomic forces, and electronic descriptors) computed at the ωB97M-D3/def2-TZVPP level of theory.

**Model Training and Evaluation Setup**

In the model evaluation experiments, we trained the models from scratch on the rMD17, MD22, and Chignolin datasets. The dataset splits, number of layers, and hidden dimensionality have been described in detail in the Results and Discussion section. For the rMD17 dataset, we used a cutoff radius of 5 Å and an initial learning rate of 0.0001. For MD22 and Chignolin, the cutoff was set to 4 Å and the initial learning rate to 0.0002. The batch size was set to 2 for both rMD17 and MD22, and 4 for Chignolin. All training procedures employed mean absolute error (MAE) as the loss function, with a loss weight of 1 for energy and 99 for forces. Early stopping was used to select the best-performing checkpoint on the validation set, which was then applied to test on all remaining molecules to generate the reported results.

For the foundation model pretraining, we employed a 6-layer equivariant neural network with a hidden dimension of 256. The batch size was set to 16 during training on the NablaDFT dataset and 8 during fine-tuning on the SPICE dataset. The learning rate was scaled based on the number of GPUs used in parallel. All training procedures were executed on eight NVIDIA GeForce RTX 4090 GPUs. Specifically, the model was trained for 100 epochs on NablaDFT and subsequently fine-tuned for 50 epochs on SPICE. A 95:5 split ratio was used for partitioning the training and validation sets during pretraining. Experimental settings for MD simulations are detailed in the Supporting Information (SI) **Methods section**.

**Data and code availability**

All relevant data supporting the key findings of this study are provided within the main article and the Supplementary Information. The benchmark datasets used in this study are publicly available, including the MD22 dataset (available at http://www.quantum-machine.org/gdml/data/npz), the rMD17 dataset (accessible at https://figshare.com/articles/dataset/Revised_MD17_dataset_rMD17_/12672038), the Chignolin dataset (available at https://github.com/microsoft/AI2BMD/tree/ViSNet/chignolin_data), the H2O dataset (available at https://github.com/BingqingCheng/ab-initio-thermodynamics-of-water). the nablaDFT dataset (available at https://github.com/AIRI-Institute/nablaDFT), and the SPICE dataset (available at https://github.com/openmm/spice-dataset).